\documentclass{article}
\PassOptionsToPackage{numbers,sort&compress}{natbib}

\usepackage[dblblindworkshop,final]{neurips_2026}
\workshoptitle{Secure and Trustworthy Quantum Machine Learning (SaTQuML)}

\usepackage[utf8]{inputenc}
\usepackage[T1]{fontenc}

\usepackage{amsmath,amssymb,amsfonts}
\usepackage{graphicx}
\usepackage{float}
\usepackage{booktabs}
\usepackage{multirow}
\usepackage{xurl}
\usepackage[hidelinks]{hyperref}
\usepackage{microtype}
\hypersetup{pdftitle={When Error Mitigation Makes Things Worse: Budget-Aware Evaluation, Extrapolation Failure, and the Calibration Trust Boundary},pdfauthor={Guilin Zhang, Kai Zhao, Xiquan Cui, Henry Heng, Xu Chu, Aletta Johanna Blanken}}

\newcommand{\ket}[1]{\left|#1\right\rangle}
\newcommand{\bra}[1]{\left\langle#1\right|}
\newcommand{\expect}[1]{\langle#1\rangle}

\newcommand{\Hideal}{\expect{H}_{\mathrm{ideal}}}
\makeatletter
\let\table@star\table
\expandafter\def\csname table*\endcsname{\table}
\expandafter\def\csname endtable*\endcsname{\endtable}
\makeatother

\title{When Error Mitigation Makes Things Worse:\\
Budget-Aware Evaluation, Extrapolation Failure,\\
and the Calibration Trust Boundary}

\author{%
  \textbf{Guilin Zhang\quad Kai Zhao\quad Xiquan Cui}\\
  \textbf{Henry Heng\quad Xu Chu\quad Aletta Johanna Blanken}\\
  Workday AI Research
}

\begin{document}
\maketitle

\begin{abstract}
Error mitigation is expected to turn noisy quantum measurements into more useful estimates. We show that it can instead amplify error. Under control-offset miscalibration, the evaluated unconstrained zero-noise extrapolation (ZNE) estimators are worse than no mitigation on 38--63\% of simulated instances and produce extreme tail errors. A small IBM Heron study finds worsening on 80--88\% of 18 instances across two depths, with mean error $3.0$--$4.1\times$ the raw error. In simulation, the median changes little, so median-only monitoring misses the failures. We then compare methods at the same online shot budget per evaluation and report offline training cost separately. After that cost is amortized, a budget-conditioned neural corrector occupies the low-budget end of the simulated accuracy--cost frontier and falls back to the raw estimate when an ensemble disagrees; its hardware transfer remains unsuccessful. Finally, we treat provider-reported calibration as an input that is not bound to the execution-time device state. A white-box projected-gradient stress test on this metadata increases the corrector's error $8.1\times$ within the feature ranges used for training and evades the disagreement monitor on half of the worsened cases. Together, the results connect budget-aware evaluation, tail risk, and calibration integrity in near-term quantum learning pipelines.
\end{abstract}

\section{Introduction}
\label{sec:intro}

Variational quantum algorithms such as the variational quantum eigensolver (VQE) and the quantum approximate optimization algorithm (QAOA) are leading candidates for near-term quantum advantage, but their expectation-value estimates are biased by device noise \citep{preskill2018quantum,cerezo2021variational,bharti2022noisy,tilly2022variational}. Error mitigation aims to remove this bias without the qubit overhead of error correction \citep{temme2017error,cai2023quantum}, and in practice it is the component that quantum machine learning pipelines trust to turn noisy hardware output into usable numbers. The most widely used technique, zero-noise extrapolation (ZNE), amplifies noise by gate folding and extrapolates the measurements back to the zero-noise limit \citep{temme2017error,li2017efficient,giurgicatiron2020digital,kim2023evidence}. Learning-based methods such as Clifford data regression (CDR) instead fit a noisy-to-ideal map on classically simulable circuits run on the same device \citep{czarnik2021error,strikis2021learning}.

This paper examines that trust from two directions that are often studied separately. The first is \emph{reliability}: does a corrected answer improve on the raw one at an affordable quantum budget, and how badly does it fail when its assumptions break? The second is \emph{security}: what information does the mitigation layer trust, and how can an adversary manipulate it?

On reliability, the usual comparison hides a first-order distinction. Methods are compared at a fixed, generous number of shots \emph{per circuit execution}, yet they need very different numbers of executions per estimate: ZNE three to five, CDR more than thirty per target circuit, a neural model trained offline exactly one. Inside a VQA loop that evaluates thousands of parameter settings under a per-iteration shot budget, these multiplicities decide which method is usable, not the accuracy at unlimited shots. \citet{bultrini2023unifying} made the budget axis explicit for classical methods; we extend it to learned mitigation and make the budget an input of the model itself.

\begin{figure}[t]
\centering
\includegraphics[width=\textwidth]{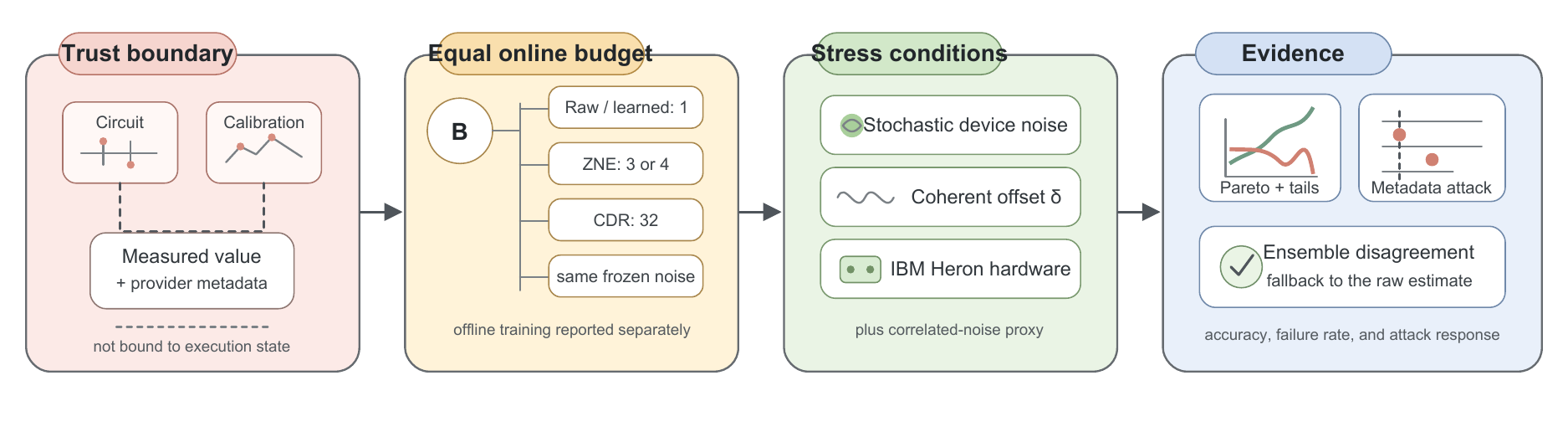}
\caption{Study design. Four mitigation pipelines share one online shot budget and a frozen noise snapshot. Three simulated regimes and a small IBM Heron run support the accuracy--cost, ZNE tail-risk, and metadata-attack analyses.}
\label{fig:study}
\end{figure}

Figure~\ref{fig:study} summarizes the study. We contribute three pieces of evidence. First, we map the tail-risk boundary of ZNE under coherent offset error and reproduce the failure pattern in a small real-hardware study. Second, we compare raw execution, learned correction, ZNE, and CDR at the same \emph{online} shot budget, while keeping the learned model's offline simulation cost visible. Third, we formalize provider-reported calibration as an execution-unbound input and evaluate a projected-gradient metadata attack. The learned corrector is a test case rather than an architectural contribution: CDR remains more accurate when its larger online budget is affordable.

\section{Related work}
\label{sec:related}

\paragraph{Classical mitigation.}
ZNE \citep{temme2017error,li2017efficient} assumes the bias varies smoothly with a controllable noise-scale factor; digital implementations amplify noise by unitary folding \citep{giurgicatiron2020digital}, with headline demonstrations at the 100-qubit scale \citep{kim2023evidence}; its reliability, however, is contingent on the extrapolation ansatz matching the actual noise response, a sensitivity we quantify in Section~\ref{sec:failure}. Probabilistic error cancellation and virtual distillation trade unbiasedness or stronger suppression for exponential sampling or additional quantum cost \citep{temme2017error,van2023probabilistic,huggins2021virtual,koczor2021exponential}. CDR \citep{czarnik2021error} fits a linear noisy-to-ideal regression on near-Clifford circuits executed on the device, paying tens of executions per target circuit but inheriting robustness to noise structure because it calibrates on the device as it actually is; \citet{strikis2021learning} extend the idea with learning-based protocols. \citet{bultrini2023unifying} show that low-cost methods win at small total shot budgets, while data-hungry methods win at large budgets. Our protocol builds on that observation.

\paragraph{Learned mitigation.}
Neural networks have been used to reconstruct observables from noisy quantum data \citep{torlai2018neural}, mitigate errors in quantum simulation \citep{bennewitz2022neural}, and predict ideal expectation values from noisy values and circuit structure \citep{liao2023machine,cantori2024synergy}. Noise-agnostic \citep{liao2025noiseagnostic} and per-circuit-adaptive \citep{adeniyi2025adaptive} variants have also been validated on small hardware experiments. Prior work rarely places learned mitigation and classical baselines on the same online shot budget, and the integrity of calibration inputs remains underexplored. We focus on those two gaps.

\paragraph{Quantum computing security.}
A growing literature attacks quantum computing at the hardware and controller layers: crosstalk-based fault injection in multi-programmed NISQ devices \citep{ashsaki2020crosstalk}, crosstalk side channels that reveal a co-tenant's circuit \citep{choudhury2025crosstalk}, and power side channels on classical controllers \citep{erata2024power}. Adversarial robustness of quantum classifiers has been studied at the model level \citep{lu2020adversarialqml}. Blind and verifiable quantum computation addresses delegated computation cryptographically \citep{broadbent2009blind,gheorghiu2019verification}, although its costs remain impractical for NISQ clouds. We study the classical mitigation layer between these settings and the execution-unbound calibration metadata it consumes.

\section{Threat model: provider calibration as a trust boundary}
\label{sec:threat}

Error mitigation is usually analyzed as a numerical problem. In a cloud pipeline it is also security-relevant because the user cannot independently verify the execution-time device state. Provider-reported calibration is authenticated as provider data, but it is not cryptographically bound to the physical process that produced a particular set of counts. We therefore treat it as a trust boundary and study how reliability failures transfer to an adversarial setting.

\paragraph{Setting.}
A mitigated pipeline spans three parties: the \emph{user}, who compiles circuits and performs classical post-processing; the \emph{provider}, who executes them on a typically multi-tenant QPU and publishes gate errors, $T_1/T_2$, readout asymmetries, and drift statistics; and \emph{co-tenants} on the same device. The user cannot independently verify the measured counts at scale or confirm that the published snapshot describes the device during execution. Blind and verifiable quantum computation \citep{broadbent2009blind,gheorghiu2019verification} address a different integrity problem and remain impractical for the NISQ setting considered here.

\paragraph{The boundary.}
Each mitigation method trusts information that the user cannot independently verify. ZNE assumes folding scales the relevant device noise, CDR assumes its near-Clifford circuits sample the same channel as the target, and a learned corrector consumes published calibration directly as features $\mathbf{x}_n$. The dependencies differ, but all sit below the user's line of visibility.

\paragraph{A1: Calibration spoofing.}
A dishonest or compromised provider, or an attacker in the metadata path, falsifies calibration while executing circuits honestly. Against NEM this is direct input poisoning. The feature ablation of Section~\ref{sec:ablations} shows that removing $\mathbf{x}_n$ triples MAE, so the controlled features are also the ones the model relies on. The sim-to-real study of Appendix~\ref{app:hardware} is a benign preview: a model trained under mischaracterized calibration overcorrects by 94\% at five layers.

\paragraph{A2: Mitigation-aware coherent bias injection.}
An adversary with access to a calibration routine or controller \citep{erata2024power} can choose a control offset $\delta$ to increase post-mitigation error. Section~\ref{sec:failure} sweeps $\delta$ over the range used by our noise model. For ZNE, mean error grows and 38--63\% of instances become worse, while the median remains near the raw baseline. A median-only dashboard would miss this tail. The small hardware run in Section~\ref{sec:hardware} shows that the same failure pattern can arise from ordinary device behavior, without deliberate injection.

\paragraph{A3: Co-tenant crosstalk adversaries.}
The device-layer literature demonstrates that a co-tenant can inject faults \citep{ashsaki2020crosstalk} and observe side channels \citep{choudhury2025crosstalk} through crosstalk on multi-programmed hardware. Our stochastic-correlated regime is an illustrative proxy for this setting, not a calibrated reproduction of a specific attack. In that proxy, ZNE is worse than the raw estimate on 44--61\% of instances, while the learned corrector improves the mean by 61--68\% (Table~\ref{tab:safety_full}).

\paragraph{Fail-safe posture and detection.}
The learned model predicts a residual correction rather than an extrapolated observable. Its outputs are not formally bounded, but the largest observed errors remain below the raw worst case in every evaluated cell. A seed-ensemble disagreement rule adds a zero-quantum-cost monitor: large disagreement returns the raw value. This is an empirical safeguard, not a robustness guarantee.
\paragraph{A PGD attack on the calibration channel.}
We test A1/A2 directly. Holding the measurement and circuit features fixed, we run projected gradient descent on the five calibration features inside an $L_\infty$ ball equal to half of each feature's training range. On miscal $n{=}8$ (150 instances), the attack raises mean error from 0.008 to 0.067 and worsens every correction (Table~\ref{tab:pgd}). At a threshold set to the 95th percentile of clean disagreement, the monitor flags half of these cases. The result is a white-box stress test of the modeled metadata range. Black-box transfer, detector-aware attacks, and empirical calibration tolerances are left to future work.

\section{Budget-conditioned neural mitigation}
\label{sec:method}

A VQA prepares $\ket{\psi(\boldsymbol{\theta})}=U(\boldsymbol{\theta})\ket{0}^{\otimes n}$ and estimates the objective $C(\boldsymbol{\theta})=\bra{\psi(\boldsymbol{\theta})}H\ket{\psi(\boldsymbol{\theta})}$. On hardware one observes an $S$-shot estimate of the noisy expectation,
\begin{equation}
\widehat{\expect{H}}_{\mathrm{noisy}} \;=\; \Hideal + \delta(\boldsymbol{\theta},\mathcal{E}) + \varepsilon_S,
\end{equation}
where $\delta$ is the bias induced by the noise process $\mathcal{E}$ and $\varepsilon_S$ is measurement noise with scale proportional to $1/\sqrt{S}$. We learn a correction function
\begin{equation}
f_\phi\!\left(\widehat{\expect{H}}_{\mathrm{noisy}},\, \mathbf{x}_c,\, \mathbf{x}_n,\, 1/\sqrt{S}\right) \;\approx\; \Hideal ,
\label{eq:model}
\end{equation}
where $\mathbf{x}_c$ contains circuit features (qubit count, layer count, parameter count, and variational angles; dimension $3+4n$) and $\mathbf{x}_n$ contains gate errors, readout asymmetries, and the calibration-drift scale when available. In deployment, $\mathbf{x}_n$ comes from the provider channel in Section~\ref{sec:threat}.

\paragraph{Architecture.}
Figure~\ref{fig:overview} (Appendix~\ref{app:diagram}) shows the model. Two multilayer-perceptron encoders embed the feature groups: the circuit encoder maps $\mathbf{x}_c$ through hidden widths $[128,256]$ to a 128-dimensional embedding, and the noise encoder maps $\mathbf{x}_n$ through width 64 to a 64-dimensional embedding. The embeddings are concatenated with the measured value and passed through an output network of widths $[256,512,256]$ (LayerNorm, GELU, dropout $0.15$) that predicts a residual correction added to the input estimate. The standard configuration has 428K parameters; Section~\ref{sec:ablations} shows accuracy is insensitive to capacity over a $5\times$ range, so the capacity is generous rather than necessary.

\paragraph{Training.}
The model minimizes Huber loss with AdamW for 75 epochs, about five minutes on a CPU workstation at the sizes studied. Training pairs $(\widehat{\expect{H}}_{\mathrm{noisy}}, \mathbf{x}_c, \mathbf{x}_n, \Hideal)$ are generated in simulation. Parameters and noise realizations are sampled per example, the input is an $S$-shot estimate, and the label is an exact noiseless value. Exact labels restrict training to classically simulable sizes.

\paragraph{Budget conditioning.}
We condition training on the shot-noise scale so that one model serves every deployment budget. Each training example draws its shot count $S$ log-uniformly from $2^{8}$ to $2^{16}$, samples the noisy input at that $S$, and appends $1/\sqrt{S}$ to $\mathbf{x}_n$. The network thereby learns how much of the input fluctuation is measurement noise versus learnable bias, so no retraining is needed across budgets.

\paragraph{Costs and break-even.}
The reported model uses 4{,}400 simulated training pairs per device family and noise regime. This offline simulation cost is not charged to the online Pareto curves. Inference adds one classical forward pass to one quantum execution. If the same training set were collected on device, its QPU cost would break even with CDR after 142 target evaluations and with ZNE after 1{,}467--2{,}200 evaluations (Appendix~\ref{app:breakeven}). Table~\ref{tab:cost} keeps this accounting explicit.

\begin{table}[t]
\caption{Per-evaluation quantum cost at total budget $B$.}
\label{tab:cost}
\centering
\footnotesize
\begin{tabular}{lccc}
\toprule
Method & Executions & Shots per execution & Offline cost \\
\midrule
Unmitigated & 1 & $B$ & --- \\
NEM (ours) & 1 & $B$ & 4.4K simulated pairs $+$ 5 min \\
ZNE (exponential) & 3 & $B/3$ & --- \\
ZNE (adaptive) & 4 & $B/4$ & --- \\
CDR & 32 & $B/32$ & repeated per target circuit \\
\bottomrule
\end{tabular}
\end{table}

\section{Experimental protocol}
\label{sec:setup}

\paragraph{Circuits and observables.}
We use hardware-efficient VQE ansatzes (alternating $R_Y,R_Z$ rotation layers and CNOT ladders, two layers) at $n=4$--$16$ qubits with the mean-magnetization observable $H=\frac{1}{n}\sum_i Z_i$. Simulations use Qiskit Aer \citep{qiskit2024}: exact density-matrix count sampling up to 10 qubits and statevector trajectory sampling above, so every reported measurement carries genuine sampling statistics. Classical baselines use mitiq~1.0 \citep{larose2022mitiq}.

\paragraph{Noise regimes.}
Three regimes span the noise-structure spectrum:
\begin{itemize}
\item \emph{Stochastic device noise}: per-instance profiles with depolarizing gate errors, thermal relaxation, and asymmetric readout. This is the regime where smooth noise scaling is most plausible.
\item \emph{Stochastic-correlated noise}: adds gate-error fluctuations and correlated $ZZ$ crosstalk as a proxy for co-tenant interference.
\item \emph{Coherent miscalibration}: models control-offset error \citep{krantz2019engineer}. In the native basis $\{R_Z,\sqrt{X},X,\mathrm{CNOT}\}$, each physical pulse receives an offset drawn once per device snapshot from $\mathcal{N}(\delta,(\delta/2)^2)$ with $\delta\in[0.02,0.06]$\,rad; virtual $R_Z$ rotations are exempt. Folded ZNE circuits and CDR training circuits face the same process. The offsets do not cancel in an inverted segment because $R_Y(\theta{+}\delta)R_Y(-\theta{+}\delta)=R_Y(2\delta)$, so error grows with the scale factor while the observable responds non-monotonically.
\end{itemize}

\paragraph{Fairness protocol.}
Each test instance freezes one device snapshot. The raw measurement, NEM input, every ZNE fold point, and every CDR training circuit are $S$-shot estimates under that snapshot; baseline failures abort the run. ZNE uses mitiq's Richardson, exponential, and adaptive-exponential factories with global folding at scales $\{1,2,3\}$; CDR uses 31 near-Clifford circuits. We report 150 instances per configuration (100 for budget sweeps), paired $t$-tests, and both means and medians because extrapolation is heavy-tailed.

\section{Results}
\label{sec:results}

\subsection{The equal-online-budget Pareto frontier}
\label{sec:pareto}

\begin{figure}[t]
\centering
\includegraphics[width=0.80\textwidth]{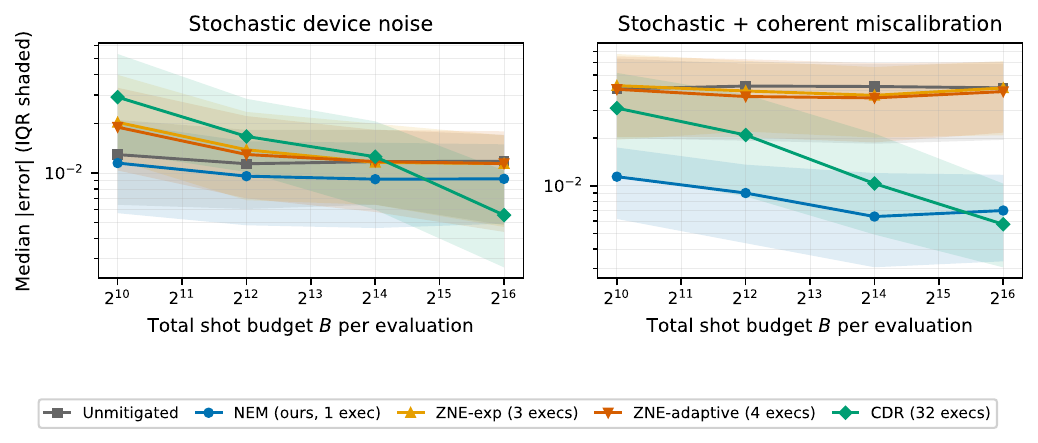}
\caption{Equal-online-budget comparison at $n=8$ (100 instances; median $|$error$|$, IQR shaded). Each method splits the same inference budget $B$ across its required executions (Table~\ref{tab:cost}); NEM's offline simulation cost is reported separately. NEM is below the raw baseline at $B\le2^{14}$ in both regimes, CDR crosses over at the largest budget, and ZNE medians track the raw line while its mean diverges.}
\label{fig:pareto}
\end{figure}

Figure~\ref{fig:pareto} compares online inference after the learned model has been trained; Table~\ref{tab:budget_full} reports every value. Under coherent miscalibration, NEM reduces mean error by 72--81\% across $B=2^{10}$--$2^{16}$. CDR rises from $+23\%$ at $B=2^{10}$ to $+84\%$ at $B=2^{16}$ and overtakes NEM at the largest budget. Both ZNE variants are negative at nearly every budget in this regime. The lone exception, adaptive ZNE at $B=2^{14}$, has a $+7\%$ mean improvement but a median near the raw baseline. Under stochastic device noise, NEM is the only positive method at $B\le2^{14}$ ($+11$ to $+20\%$), while CDR turns positive at $B=2^{16}$ ($+47\%$). These are online comparisons; Section~\ref{sec:method} and Appendix~\ref{app:breakeven} report the offline training cost.

The practical advantage therefore depends on reuse. A learned corrector is attractive when one trained model serves many parameter settings; CDR is preferable when its larger per-target budget is affordable or reuse is limited.

\subsection{When extrapolation fails, and how badly}
\label{sec:failure}

\begin{figure}[t]
\centering
\includegraphics[width=0.55\textwidth]{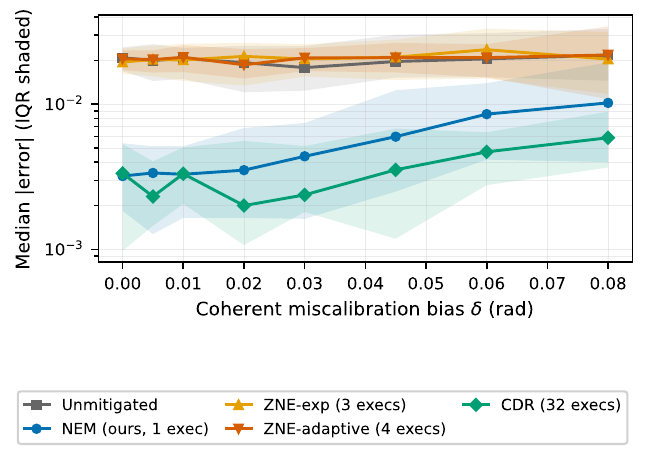}
\caption{Median $|$error$|$ versus miscalibration strength $\delta$ at $n=8$ (80 instances per point, 8{,}192 shots per execution). ZNE medians coincide with the unmitigated baseline at every $\delta$; NEM matches CDR at small $\delta$ at $1/32$ the quantum cost and declines smoothly as the unlearnable coherent share grows.}
\label{fig:boundary}
\end{figure}

Figure~\ref{fig:boundary} sweeps $\delta$ from 0 to 0.08\,rad over a fixed stochastic background. ZNE's median improvement never exceeds $+2\%$, including at $\delta=0$. The remaining bias there is dominated by readout error, which gate folding does not amplify.

Mean error grows sharply with $\delta$; exponential extrapolation reaches $-646\%$ improvement at $\delta=0.06$. ZNE is worse than the raw estimate on 38--63\% of instances. Its extrapolated output is unconstrained: one $n{=}12$ instance reaches error 371 in observable units, three orders of magnitude above the largest raw error in that cell. The median-preserving, tail-growing response also defines the attack surface for A2.

\paragraph{Physical-range safeguards.}
For our normalized observable, the ideal value lies in $[-1,1]$. Clipping an estimate $a$ to $\Pi_{[-1,1]}(a)$ cannot increase its absolute error against such a target and caps that error at 2. Thus the error of 371 diagnoses an unconstrained estimator, not an unavoidable property of ZNE. The tables and curves use the original unconstrained fits; neither clipped estimates nor physically constrained fits were evaluated. Their failure rates and mean errors cannot be inferred from these summaries. Appendix~\ref{app:clipping} states the bound and the comparison needed to assess these safeguards.

NEM, by contrast, tracks CDR at small $\delta$ ($+82\%$ versus $+83\%$ mean at $\delta=0$) at $1/32$ of the quantum cost. Its improvement then declines smoothly to $+44\%$ at $\delta=0.08$, as the high-dimensional oscillatory component of the coherent bias exceeds what the network can learn from 4{,}000 samples. CDR, which re-fits per instance, stays at 69--86\%.

\begin{table}[t]
\caption{Robustness statistics, coherent-miscalibration regime, $n{=}10$ (150 instances, 8{,}192 shots per execution). ``Worse'': fraction of instances whose error exceeds the unmitigated error.}
\label{tab:safety}
\centering
\footnotesize
\begin{tabular}{lccccc}
\toprule
Method & Mean & Median & p95 & Max & Worse \\
\midrule
Unmitigated & .0412 & .0385 & .080 & 0.11 & --- \\
NEM (ours) & \textbf{.0061} & \textbf{.0044} & \textbf{.015} & \textbf{0.03} & 7\% \\
ZNE-Richardson & .0460 & .0432 & .099 & 0.15 & 63\% \\
ZNE-exponential & .1569 & .0385 & .081 & 17.1 & 49\% \\
ZNE-adaptive & .1220 & .0394 & .082 & 12.1 & 53\% \\
CDR ($32\times$ cost) & .0035 & .0025 & .009 & 0.01 & 6\% \\
\bottomrule
\end{tabular}
\end{table}

The asymmetry in failure behavior matters as much as the averages. Across all 2{,}000 structured-regime test instances (4--20 qubits), the largest observed neural-corrector error is below the largest raw error in every cell. This is an empirical result, not a formal output bound.

A safeguard tightens the observed result. We train three seeds and return the raw value when they disagree beyond a threshold calibrated on unlabeled data. In the representative cell, this lowers the worse-than-raw rate from 7\% to 5\% at a cost of three percentage points of mean improvement. Section~\ref{sec:threat} evaluates the same rule under attack.

\subsection{Real-hardware validation}
\label{sec:hardware}

We next test whether the simulated failure pattern appears on a real device. The experiment uses hardware-efficient VQE on \texttt{ibm\_marrakesh}, a 156-qubit Heron~r2 processor, with four qubits and two circuit depths. We compare raw execution with folding ZNE under the equal-online-budget protocol using ten and eight instances, 8{,}192 shots, and one batch. No additional noise is injected.

\begin{figure}[t]
\centering
\includegraphics[width=0.58\textwidth]{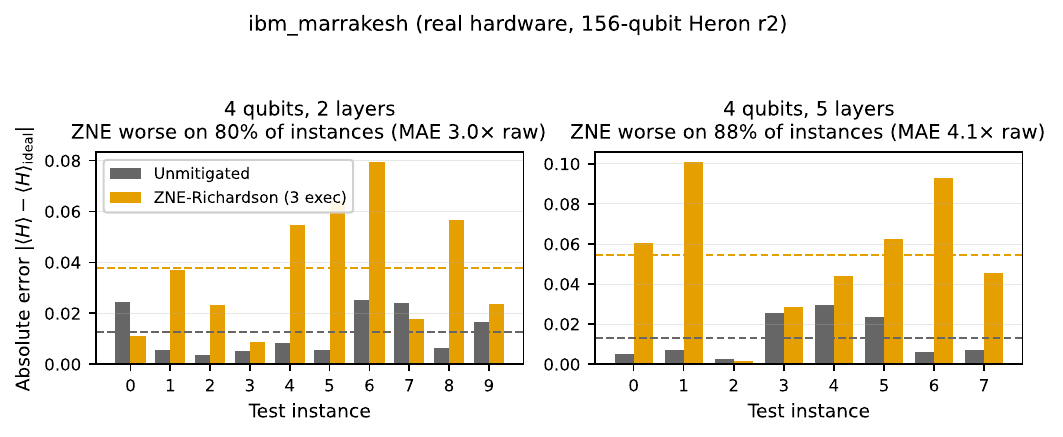}
\caption{Folding ZNE on real hardware (\texttt{ibm\_marrakesh}, 4 qubits): per-instance absolute error for the raw estimate (grey) and ZNE-Richardson (orange); dashed lines are means. In this one-batch study, ZNE is worse on 80\% (2 layers, 10 instances) and 88\% (5 layers, 8 instances), with mean error $3.0\times$/$4.1\times$ above the raw level.}
\label{fig:hardware}
\end{figure}

Figure~\ref{fig:hardware} shows the result. Mean ZNE error is $3.0\times$ the raw error at two layers and $4.1\times$ at five layers; ZNE is worse on 80\% and 88\% of instances, respectively. This small study reproduces the direction and tail pattern seen in simulation. It does not isolate control offset from other hardware effects, and its sample size is too small for a broad hardware claim. Repeated calibrations, qubit subsets, and devices are left to future work.

This run is not a hardware-accuracy result for the learned corrector. A twin-trained model does not transfer to these near-shot-noise-floor circuits, as detailed in Appendix~\ref{app:hardware}. The negative transfer result limits the deployment claim and motivates on-device adaptation \citep{liao2025noiseagnostic,adeniyi2025adaptive}.

\subsection{Scaling, controls, and the model's attack surface}
\label{sec:scaling}

Separately trained models maintain their simulated improvement from 4 to 20 qubits (Figure~\ref{fig:scaling} and Table~\ref{tab:scaling_full}). Mean error reduction is 72--85\% under coherent miscalibration and 61--68\% in the stochastic-correlated proxy. This is a per-size scaling study, not cross-size generalization. At $n{=}16$ and $n{=}20$, NEM matches CDR under miscalibration at $1/32$ of CDR's online executions. Under purely stochastic noise, improvement is smaller ($+7$ to $+22\%$) as the raw estimate approaches the shot-noise floor.

\label{sec:ablations}
\paragraph{Direct prediction.}
Is the network mitigating, or memorizing the expectation-value landscape? A control with identical architecture but no access to the noisy measurement answers this: it is worse than unmitigated execution by $57$--$927\%$ at every size and regime (Table~\ref{tab:scaling_full}). The network's role is correction, not regression of $\Hideal$ from circuit parameters.

\paragraph{Features.}
Table~\ref{tab:ablation} isolates what the network uses at $n=8$. Removing device features triples MAE, while removing circuit features changes it by only $0.3\%$. With both groups removed, the measurement-conditioned model still improves over the raw estimate by 46\%. The same dependence defines the A1 attack surface. Capacity is not decisive: an 88K-parameter model matches the 428K model ($+81.4\%$ versus $+81.7\%$).

\section{Discussion and limitations}
\label{sec:discussion}

\paragraph{Claims and non-claims.}
At the same online budget, single-execution learned correction occupies the low-budget end of the accuracy--cost frontier after its offline cost is amortized. CDR is more accurate when its 32 executions per target are affordable. The network is a plain residual MLP, and its favorable worst-case behavior is empirical rather than certified. Our security experiment is likewise scoped: it is a white-box metadata stress test within the modeled feature range, not evidence that current cloud calibration APIs have been compromised.

\paragraph{Limitations.}
The evaluation is primarily in simulation. The hardware study has only ten and eight instances from one batch, and the twin-trained corrector does not transfer without on-device adaptation (Appendix~\ref{app:hardware}). The noise models are not device-fitted, observables are diagonal, each model is trained for one device family and regime, and the ansatz family is fixed. We also report paired $t$-tests from the original evaluation; future work should add repeated hardware calibrations and nonparametric confidence intervals for heavy-tailed cells.

The metadata budget is not calibrated to observed device drift: an in-range perturbation need not be small or physically realizable. Matched-budget random perturbations and drift-calibrated budgets remain unevaluated. Best-validation-seed selection also leaves the stability of gains across all training seeds unresolved. If changing calibration requires retraining, the amortization period restarts and adaptation costs must be added before claiming a deployment advantage.

\section{Conclusion}
\label{sec:conclusion}

Mitigation methods should be compared under a fixed online budget and with their offline assumptions visible. Under that accounting, ZNE can amplify tail error while leaving the median nearly unchanged. A learned corrector can be useful after training is amortized, but its hardware transfer remains unresolved. Treating provider calibration as execution-unbound input also exposes a concrete metadata attack surface. These results motivate evaluations that report cost, tail risk, and calibration integrity together.

\clearpage
{\small
\bibliographystyle{plainnat}
\bibliography{references}
}

\clearpage
\appendix

\section{Complete result tables}
\label{app:tables}

\begin{figure}[H]
\centering
\includegraphics[width=0.70\textwidth]{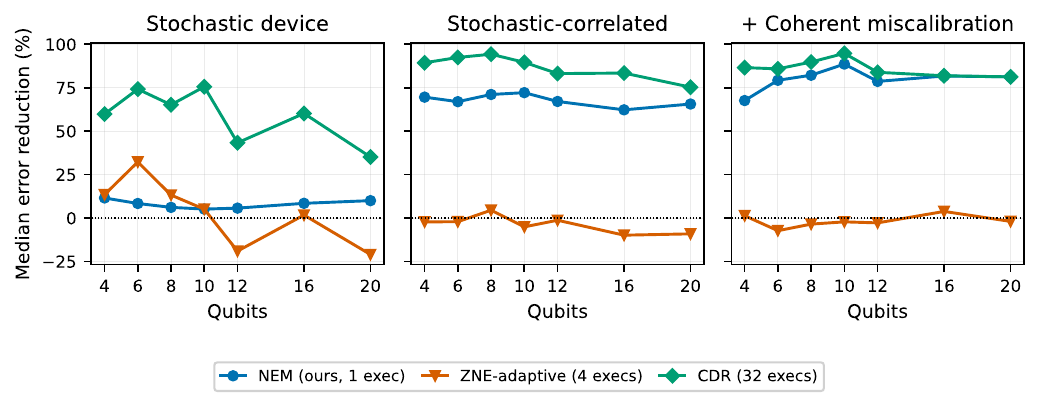}
\caption{Median error reduction versus qubit count (150 instances per cell, 100 at $n{=}20$; $S{=}8{,}192$ shots per execution up to $n{=}10$; $1{,}024$ at $n{=}12,16$; $512$ at $n{=}20$). NEM's improvement does not degrade from 4 to 20 qubits in the structured regimes; the direct-prediction control (omitted for scale) is negative at every size.}
\label{fig:scaling}
\end{figure}

\begin{table}[H]
\caption{Mean error reduction vs.\ unmitigated execution (\%) for every cell of the scaling study. 150 test instances per cell (100 at $n{=}20$); per-execution shots $S{=}8192$ ($n\le10$), $1024$ ($n{=}12,16$), $512$ ($n{=}20$). Asterisks mark cells where the NEM improvement is not significant at $p<0.05$ (paired $t$-test).}
\label{tab:scaling_full}
\centering\small
\resizebox{\textwidth}{!}{%
\begin{tabular}{llccccccc}
\toprule
Regime & Method & $n{=}4$ & $n{=}6$ & $n{=}8$ & $n{=}10$ & $n{=}12$ & $n{=}16$ & $n{=}20$ \\
\midrule
Stochastic device & NEM (ours) & $+22$ & $+15$ & $+10$ & $+13$ & $+11$ & $+7$$^{*}$ & $+11$ \\
 & Direct prediction & $-927$ & $-631$ & $-605$ & $-395$ & $-365$ & $-226$ & $-180$ \\
 & ZNE-Richardson & $-32$ & $-34$ & $-46$ & $-29$ & $-160$ & $-102$ & $-144$ \\
 & ZNE-exponential & $+21$ & $-87$ & $-35$ & $+7$ & $-894$ & $-206$ & $-1152$ \\
 & ZNE-adaptive & $+20$ & $+21$ & $+15$ & $<-10^4$ & $-9970$ & $-6546$ & $<-10^4$ \\
 & CDR & $+65$ & $+73$ & $+74$ & $+78$ & $+44$ & $+55$ & $+46$ \\
\midrule
Stochastic-correlated & NEM (ours) & $+68$ & $+66$ & $+66$ & $+67$ & $+64$ & $+62$ & $+61$ \\
 & Direct prediction & $-302$ & $-186$ & $-156$ & $-120$ & $-74$ & $-57$ & $-7$ \\
 & ZNE-Richardson & $-8$ & $-8$ & $+6$ & $-3$ & $-38$ & $-33$ & $-42$ \\
 & ZNE-exponential & $-188$ & $+3$ & $-15$ & $-27$ & $-38$ & $-120$ & $-166$ \\
 & ZNE-adaptive & $<-10^4$ & $+8$ & $-839$ & $-198$ & $<-10^4$ & $-2672$ & $-8$ \\
 & CDR & $+90$ & $+90$ & $+93$ & $+88$ & $+78$ & $+81$ & $+69$ \\
\midrule
Coherent miscalibration & NEM (ours) & $+72$ & $+77$ & $+80$ & $+85$ & $+76$ & $+80$ & $+77$ \\
 & Direct prediction & $-244$ & $-188$ & $-129$ & $-57$ & $-69$ & $-3$ & $-0$ \\
 & ZNE-Richardson & $-29$ & $-11$ & $-10$ & $-12$ & $-17$ & $-6$ & $-29$ \\
 & ZNE-exponential & $-104$ & $-24$ & $-70$ & $-281$ & $-87$ & $-464$ & $-32$ \\
 & ZNE-adaptive & $+12$ & $<-10^4$ & $-6204$ & $-196$ & $-6250$ & $<-10^4$ & $-6604$ \\
 & CDR & $+85$ & $+83$ & $+88$ & $+92$ & $+79$ & $+81$ & $+79$ \\
\bottomrule
\end{tabular}
}
\end{table}

\begin{table}[H]
\caption{Equal-budget sweep at $n{=}8$ (100 instances): mean error reduction vs.\ unmitigated execution (\%) when every method spends the same total budget $B$ per evaluation point, split across its required executions.}
\label{tab:budget_full}
\centering\small
\begin{tabular}{llcccc}
\toprule
Regime & Method & $B{=}2^{10}$ & $B{=}2^{12}$ & $B{=}2^{14}$ & $B{=}2^{16}$ \\
\midrule
Stochastic device & NEM (ours) & $+11$ & $+17$ & $+20$ & $+19$ \\
 & ZNE-exponential & $-194$ & $-216$ & $-6$ & $+8$ \\
 & ZNE-adaptive & $<-10^4$ & $<-10^4$ & $<-10^4$ & $+10$ \\
 & CDR & $-132$ & $-54$ & $-9$ & $+47$ \\
\midrule
Coherent miscalibration & NEM (ours) & $+72$ & $+77$ & $+81$ & $+81$ \\
 & ZNE-exponential & $-471$ & $-36$ & $-93$ & $-129$ \\
 & ZNE-adaptive & $<-10^4$ & $-183$ & $+7$ & $-76$ \\
 & CDR & $+23$ & $+42$ & $+68$ & $+84$ \\
\bottomrule
\end{tabular}
\end{table}

\begin{table}[H]
\caption{Tail statistics across structured-regime cells: median and maximum absolute error, and the fraction of instances on which each method returns a worse answer than unmitigated execution.}
\label{tab:safety_full}
\centering\small
\resizebox{\textwidth}{!}{%
\begin{tabular}{llccc|ccc|ccc}
\toprule
 & & \multicolumn{3}{c|}{NEM (ours)} & \multicolumn{3}{c|}{ZNE (best of three)} & \multicolumn{3}{c}{CDR} \\
Cell & Raw MAE & med & max & worse & med & max & worse & med & max & worse \\
\midrule
Coherent miscal., $n{=}10$ & 0.0412 & 0.0044 & 0.03 & 7\% & 0.0432 & 0.15 & 63\% & 0.0025 & 0.01 & 6\% \\
Coherent miscal., $n{=}12$ & 0.0397 & 0.0080 & 0.03 & 9\% & 0.0399 & 0.17 & 59\% & 0.0073 & 0.02 & 6\% \\
Coherent miscal., $n{=}16$ & 0.0442 & 0.0079 & 0.03 & 11\% & 0.0419 & 0.14 & 49\% & 0.0078 & 0.02 & 10\% \\
Coherent miscal., $n{=}20$ & 0.0436 & 0.0082 & 0.03 & 12\% & 0.0504 & 0.19 & 61\% & 0.0087 & 0.03 & 7\% \\
Coherent miscal., $n{=}4$ & 0.0436 & 0.0115 & 0.04 & 14\% & 0.0349 & 0.16 & 38\% & 0.0046 & 0.02 & 4\% \\
Coherent miscal., $n{=}6$ & 0.0390 & 0.0072 & 0.04 & 12\% & 0.0388 & 0.12 & 57\% & 0.0047 & 0.02 & 16\% \\
Coherent miscal., $n{=}8$ & 0.0437 & 0.0074 & 0.03 & 10\% & 0.0430 & 0.14 & 54\% & 0.0043 & 0.01 & 4\% \\
Stochastic-corr., $n{=}10$ & 0.0345 & 0.0090 & 0.04 & 11\% & 0.0346 & 0.11 & 50\% & 0.0037 & 0.01 & 4\% \\
Stochastic-corr., $n{=}12$ & 0.0348 & 0.0105 & 0.07 & 13\% & 0.0350 & 1.82 & 49\% & 0.0055 & 0.02 & 10\% \\
Stochastic-corr., $n{=}16$ & 0.0349 & 0.0128 & 0.06 & 17\% & 0.0401 & 0.15 & 61\% & 0.0059 & 0.02 & 6\% \\
Stochastic-corr., $n{=}20$ & 0.0360 & 0.0119 & 0.07 & 19\% & 0.0377 & 0.10 & 52\% & 0.0092 & 0.04 & 10\% \\
Stochastic-corr., $n{=}4$ & 0.0395 & 0.0108 & 0.04 & 15\% & 0.0360 & 0.11 & 55\% & 0.0045 & 0.01 & 8\% \\
Stochastic-corr., $n{=}6$ & 0.0386 & 0.0115 & 0.04 & 15\% & 0.0355 & 0.08 & 44\% & 0.0030 & 0.01 & 0\% \\
Stochastic-corr., $n{=}8$ & 0.0389 & 0.0109 & 0.06 & 16\% & 0.0340 & 0.17 & 46\% & 0.0023 & 0.01 & 2\% \\
\bottomrule
\end{tabular}
}
\end{table}

\begin{table}[H]
\caption{PGD attack on the calibration channel (miscal $n{=}8$, 150 instances, $L_\infty$ budget $=$ half each feature's training range). The clean model is reconstructed to $+81\%$ improvement; the attack perturbs only the published calibration $\mathbf{x}_n$, not the measurement. ``Stealthy success'': fraction of worsened instances the disagreement monitor fails to flag.}
\label{tab:pgd}
\centering
\footnotesize
\begin{tabular}{lccccc}
\toprule
Clean err & Attacked err & Inflation & Worsened & Detection & Stealthy success \\
\midrule
.0082 & .0671 & $8.1\times$ & 100\% & 50\% & 50\% \\
\bottomrule
\end{tabular}
\end{table}

\begin{table}[H]
\caption{Feature ablation, coherent-miscalibration regime, $n{=}8$ (150 instances; unmitigated MAE $.0437$).}
\label{tab:ablation}
\centering
\small
\begin{tabular}{lccc}
\toprule
Variant & MAE & vs.\ raw & vs.\ full \\
\midrule
Full model & .0081 & $+81\%$ & --- \\
No circuit features & .0081 & $+81\%$ & $+0.3\%$ \\
No noise features & .0236 & $+46\%$ & $+191\%$ \\
Noisy value only & .0237 & $+46\%$ & $+192\%$ \\
Direct prediction (no measurement) & .0995 & $-129\%$ & --- \\
\bottomrule
\end{tabular}
\end{table}

\section{Break-even analysis}
\label{app:breakeven}

Our experiments generate the 4{,}400 training pairs in simulation, so the main online curves do not treat them as QPU calls. For a deployment that collects an equivalent training set on device, let $M$ be the number of target circuits served before retraining. At equal shots per execution, the conditional QPU cost is $4400+M$ for NEM, $kM$ for ZNE ($k=3$ or $4$), and $32M$ for CDR. The corresponding break-even points are 142 targets relative to CDR, 1{,}467 relative to adaptive ZNE, and 2{,}200 relative to exponential ZNE. Table~\ref{tab:breakeven} reports this conditional accounting; it should not be read as the cost of the simulation-trained model used in our experiments.

\begin{table}[H]
\caption{Break-even: distinct evaluation points $M^\star$ beyond which NEM's total quantum cost (offline plus one execution per point) falls below the baseline's.}
\label{tab:breakeven}
\centering
\small
\begin{tabular}{lcc}
\toprule
NEM vs.\ & Per-eval.\ executions & Break-even $M^\star$ \\
\midrule
ZNE (exponential) & 3 & 2{,}200 \\
ZNE (adaptive) & 4 & 1{,}467 \\
CDR & 32 & 142 \\
\bottomrule
\end{tabular}
\end{table}

\section{Hardware sim-to-real study}
\label{app:hardware}

Section~\ref{sec:hardware} reports the ZNE pattern on \texttt{ibm\_marrakesh} and notes that a twin-trained corrector does not transfer. We give the full result here because the negative transfer identifies a limit of simulation-trained mitigation and a benign analogue of A1.

The sim-to-real protocol reads the runtime \texttt{Target}, builds an Aer noise model with \texttt{NoiseModel.from\_backend}, trains on that digital twin, and tests on the processor. No device data enters training. The twin overstates the observed bias, so the learned correction is too aggressive: improvement is $-3\%$ at two layers and $-94\%$ at five layers.

\begin{table}[H]
\caption{Hardware results on \texttt{ibm\_marrakesh} (4 qubits, equal-budget, $8{,}192$ shots). Improvement is mean error reduction vs.\ unmitigated; ``worse'' is the fraction of instances exceeding the unmitigated error. ZNE's failure transfers to hardware; the twin-trained neural model's correction does not.}
\label{tab:hardware}
\centering
\small
\begin{tabular}{llccc}
\toprule
Depth & Method & Improvement & Worse & MAE \\
\midrule
\multirow{2}{*}{2 layers} & ZNE-Richardson & $-201\%$ & 80\% & .038 \\
 & NEM (twin-trained) & $-3\%$ & 60\% & .013 \\
\midrule
\multirow{2}{*}{5 layers} & ZNE-Richardson & $-314\%$ & 88\% & .055 \\
 & NEM (twin-trained) & $-94\%$ & 88\% & .026 \\
\bottomrule
\end{tabular}
\end{table}

The hardware result exposes a distribution shift between the digital twin and the physical device. In simulation, training and test share one noise process; on hardware, the twin overstates the observed bias and the model overcorrects. The ensemble rule can return the raw estimate when seeds disagree, but it does not certify behavior under this shift. Closing the gap requires on-device data, label-free augmentation \citep{liao2025noiseagnostic}, or adaptive per-circuit conditioning \citep{adeniyi2025adaptive}.

\section{Model diagram}
\label{app:diagram}

\begin{figure}[H]
\centering
\includegraphics[width=0.9\textwidth]{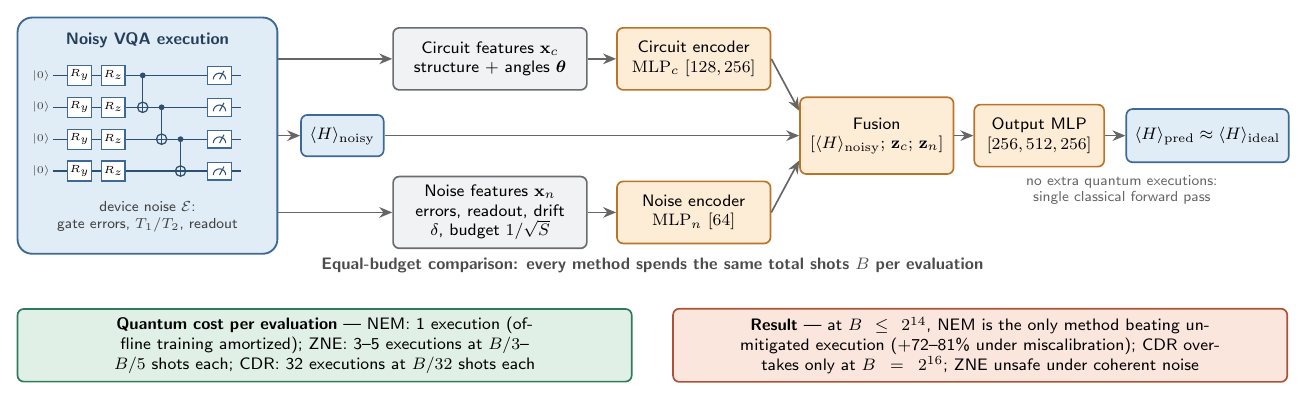}
\caption{Budget-conditioned neural error mitigation. A single $S$-shot noisy VQA execution is combined with circuit and device embeddings, including the shot-noise scale $1/\sqrt{S}$, to predict the ideal expectation value. Under an equal online budget $B$, competing methods split $B$ across their 3--32 required executions.}
\label{fig:overview}
\end{figure}

\section{Reproducibility details}
\label{app:repro}

\paragraph{Data generation.} Each scaling cell uses 4{,}000 training and 400 validation pairs (2{,}500/250 at $n{=}20$), with circuit parameters drawn uniformly from $[0,2\pi)$ and noise parameters drawn per sample from the regime ranges of Section~\ref{sec:setup}. Noisy inputs are measured estimates at the per-size shot budget; ideal labels are exact statevector expectation values.

\paragraph{Training.} AdamW, learning rate $10^{-3}$, weight decay $10^{-4}$, batch size 64, Huber loss, cosine annealing with warm restarts ($T_0{=}25$), 75 epochs, gradient clipping at norm 1. The best-validation checkpoint is selected. Three seeds are trained per cell; the best-validation seed is reported and the seed ensemble feeds the safeguard analysis.

\paragraph{Baselines.} mitiq 1.0 with global unitary folding at scale factors $\{1,2,3\}$ (adaptive: four steps, scale factor 2, depolarizing asymptote). CDR: 31 near-Clifford training circuits, fraction of non-Clifford gates retained $0.5$, target circuits transpiled to the $\{R_Z,\sqrt{X},X,\mathrm{CNOT}\}$ basis with an exact expectation-equivalence check. Every baseline execution passes through the same measured-shot estimator and, in the miscalibration regime, the same per-snapshot circuit transform as the raw measurement.

\paragraph{Statistics.} Improvements are reported against the unmitigated estimate on the same instances; significance uses two-sided paired $t$-tests on absolute errors. Medians and interquartile ranges are reported wherever extrapolation tails dominate means.

\section{Physical-range clipping and the limits of aggregate results}
\label{app:clipping}

Let $t\in[-1,1]$ be an ideal expectation and let $a\in\mathbb{R}$ be an extrapolated value. Define $c(a)=\max(-1,\min(1,a))$. Projection onto an interval satisfies
\begin{equation}
|c(a)-t|\leq |a-t|,\qquad |c(a)-t|\leq 2.
\end{equation}
If $a$ is inside the interval, the error is unchanged. If $a>1$, then $|c(a)-t|=1-t\leq a-t$; the case $a<-1$ is analogous. The same argument applies to any learned or regression-based estimate of this observable.

This deterministic safeguard eliminates unbounded numerical tails, but does not guarantee improvement over the raw estimate. For example, $t=0$, a raw estimate of $0.1$, and $a=3$ give clipped error 1, still larger than the raw error $0.1$. Conversely, some originally worsened cases may cease to be worsened after clipping. The clipped worse-than-raw fraction cannot exceed the original fraction on the same instances, but its actual value requires the paired ideal, raw, and extrapolated outputs.

Our reported aggregate tables are insufficient to reconstruct those paired outputs. Consequently, this version provides the bound, not a new empirical clipping result. A follow-up comparison should apply physical-range projection to every estimator, evaluate constrained extrapolation separately, and report paired mean, median, quantiles, and worse-than-raw rates with bootstrap intervals. Constrained fitting changes the fit itself and is not equivalent to clipping its final output. The original paired $t$-tests do not establish that the conclusions would persist under either modification.

\end{document}